# Adaptive Intrusion Detection System Leveraging Dynamic Neural Models with Adversarial Learning for 5G/6G Networks

Neha
University of Newcastle
Callaghan NSW, Australia
neha10@uon.edu.au

Tarunpreet Bhatia*
Department of Computer Science and Engineering
Thapar Institute of Engineering and Technology
Patiala, India
tarunpreet@thapar.edu

*Abstract*— Intrusion Detection Systems (IDS) are critical components in safeguarding 5G/6G networks from both internal and external cyber threats. While traditional IDS approaches rely heavily on signature-based methods, they struggle to detect novel and evolving attacks. This paper presents an advanced IDS framework that leverages adversarial training and dynamic neural networks in 5G/6G networks to enhance network security by providing robust, real-time threat detection and response capabilities. Unlike conventional models, which require costly retraining to update knowledge, the proposed framework integrates incremental learning algorithms, reducing the need for frequent retraining. Adversarial training is used to fortify the IDS against poisoned data. By using fewer features and incorporating statistical properties, the system can efficiently detect potential threats. Extensive evaluations using the NSL-KDD dataset demonstrate that the proposed approach provides better accuracy of 82.33% for multiclass classification of various network attacks while resisting dataset poisoning. This research highlights the potential of adversarial-trained, dynamic neural networks for building resilient IDS solutions.

## I. Introduction

5G/6G networks enable ultra-low latency, massive device connectivity, and critical applications such as IoT, smart cities, autonomous vehicles, and remote healthcare. This vast and diverse connectivity landscape expands the attack surface, exposing networks to sophisticated threats like Distributed Denial of Service (DDoS), advanced persistent threats, and network slicing attacks. An Intrusion Detection System (IDS) plays a crucial role by continuously monitoring network traffic, detecting anomalies, and identifying malicious activities in real-time by any category of user/software (e.g., the computing systems [1, 2], the sensors [3, 4] and the software-defined networks [5]). Its ability to adapt and learn from evolving threats ensures proactive threat detection and response. Furthermore, as 5G/6G networks prioritize reliability, security, and privacy, an IDS becomes indispensable for maintaining secure communication, minimizing downtime, protecting user data, and enabling safe adoption of next-generation applications and services.

An IDS is generally classified into two types based on the methodologies employed for detecting intrusions i.e. signature-based and anomaly-based intrusion detection systems. A signature-based IDS detects attacks by looking for pre-defined and pre-determined patterns, such as byte sequences in network traffic or known malicious instruction flows used by malware. An anomaly-based IDS detects both network and system intrusions by observing the system activity and classifying it as either normal or abnormal using statistical, and machine learning approaches. An IDS using a signature-based approach uses rule-based matching [6]. An IDS using an anomaly-based system applies supervised or semi-supervised machine learning algorithms to train the system with a collection of labelled and unlabelled network data. Previous literature has reported that anomaly-based IDS has a high detection rate and fast processing speed [7]. With the increase in network traffic and complex heterogeneity of the network attacks, anomaly-based IDS has gained significance. The detection accuracy and cost of the anomaly-based IDS [8, 9] significantly depend on relevant network data features extraction and the pertinent classification model employed. Feature redundancy and data imbalance in the training set can result in a bottleneck in developing real IDS [10, 11].

In the network security community, anomaly-based IDS adapting deep learning methodologies has shown better accuracy because hierarchical structured deep neural networks extract representative features of the data. But when we deal with complex systems, there is a fair possibility that methods cannot measure all state components. Dynamic neural nets can successfully handle this kind of input and demonstrate functional behavior in the presence of unmodeled dynamics because they can cooperate with various input structures. The first dynamic neural nets have been proposed by Hopfield [12] and later analyzed in [13]. This paper will introduce a dynamic neural net classifier to handle the input's unmodeled dynamics and strengthen the model's detection against a specific set of samples.

Data scarcity and data imbalance also can formulate challenges such as data poisoning and altering the statistical property of the data encountered in anomaly-based IDS. One of the solutions could be to increase the number of related data samples in the training set. However, labeling a large dataset can be expensive, time-consuming, and infeasible with increased intrusion attacks. Thus, we propose a data augmentation module that promotes adversarial learning and statistical learning techniques to prevent feeding models with unrealistic data and rather populate it with statistically similar data using some conditional generators.

In an anomaly-based IDS, the formulation of a machine-learning model does not involve an explicit mechanism. There has often been an implicit assumption that the training set is available before the training process, and the learning ceases post-training process. With the increase in cyber-attacks and cyber threats, the induced classification model can only detect intrusions based on training data provided at the early stage. Although IDS is utilized to detect a wide range of cyber-attacks and threats, this approach to learning is not fruitful. Ultimately, the model is worn out and gets polluted with concept drifts and stale data with no information regarding the latest attack footprints. It could only be tackled by retraining the model with new data on periodic intervals to cope with new attack patterns. However, the cost of retraining the model with an increasing stream of data is not efficient and is also a time-consuming process. Thus, we propose to use incremental learning techniques, in which the model is continuously updated and extended without retraining and using heavy computational resources. It ensures that the model can adapt to the new data without forgetting the existing knowledge [14]. By employing adversarial learning, the IDS can simulate and learn from potential attack scenarios, making it more resilient to sophisticated cyber threats and reducing false positives. This is particularly critical in 5G/6G environments, which demand low latency, high-speed communication, and secure connectivity for diverse and complex applications.

## II. Related Work

There have been various attempts to showcase and use machine learning in the network security field. However, network security slightly drifts from other areas in terms of dynamism and data status that keeps on changing in real-time. In the category of machine learning-based IDS, Zhang et al. [11] have proposed a combination of unsupervised clustering and supervised learning for classifying network traffic. Ashfaq et al. [15] proposed a semi-supervised fuzzy method for intrusion detection. These models are trained with a huge amount of network traffic data. However, the detection accuracy of the models is unclear for a small amount of network data. Apart from these methodologies, supervised classification models such as SVM, random forest, Logistic Regression etc. have been widely applied to improve the performance of modern IDS [10, 16-18]. Intrusion detection can be achieved through various techniques; however, the challenge arises due to performance constraints and the presence of class imbalance within the dataset, complicating the detection process. Ambusaidi et al. [10] introduced a mutual information-based algorithm to select network features for handling feature redundancy problems. RF and tree algorithms are employed to ensemble the sub-classification model to mitigate the overfitting problem. Louk and Tama [8] proposed a dual ensemble model for IDS combining bagging and gradient boosting decision tree (GBDT) techniques to improve detection rates and reduce false alarms. Alem et al. [9] combined two types of IDSs with a neural network-based decision-making system and validated in a real industrial environment. However, the models have not mitigated issues such as malicious data attacks and emerging drifts in the statistical properties of the data in an IDS. The state-of-the-art models do not have any mechanism to extend the existing model's knowledge with manageable effort.

In deep learning-based IDSs, Tang et al. [5] applied three-layer DNNs for extracting multilevel features and classifying flow-based network intrusions. Yin et al. [19] proposed an IDS system that employs a recurrent neural network with a fine-tuned number of hidden nodes and learning parameters. Kim et al. [20] applied four hidden layers, 100 hidden units, and the ReLU function as an activation function of the hidden layer to decrease the false alarm rate of the IDS. However, challenges such as poisoning attacks and altering the statistical properties of the data have become the bottleneck of developing an accurate IDS.

The authors employ Generative Adversarial Networks (GANs) to enhance the performance of their IDS by generating synthetic network traffic, thereby addressing issues of class imbalance and improving the model's ability to detect novel attack patterns [21-25]. However, the question is about the diversity of the output generated by GANs and the model collapse. Zhang et al. [21] proposed an IDS system that adopts augmented intrusion data to train supervised models for detecting network intrusions. The model handles the data imbalance and data scarcity problem using data augmentation methods, namely, probabilistic generative models and GANs. Xiong et al. [22] enhanced the robustness of IDS by using adversarial training to improve detection accuracy. Zacaron et al. [23] proposed anomaly-based IDS for Software-defined Networking (SDN), leveraging GAN, deep convolutional GAN, and Wasserstein GAN with Gradient Penalty (WGAN-GP). However, the cost of training the model with an increasing stream of data is an inefficient and time-consuming process.

## III. Threat Model

Network Intrusion attacks aim at infringing the intrusion data pattern present in the dataset. The goal of these attacks is to make the model inconsistent and unreliable in the target environment. One of the common attacks is a data poisoning attack. Data Poisoning is an adversarial attack that attempts to manipulate the training dataset to control the prediction behavior of a trained model (as shown in Fig 1). Our threat model simulates the same attack where it will flip the labels for random records with the other known labels. Previous work has investigated attacks and defenses for data poisoning attacks applied to PCA [26], feature selection [27], linear regression [28], and SVMs [29], to name a few. Limited-knowledge attacks simulate a more pragmatic attack environment, where the training data or learning algorithms are not known. In this environment, an adversary uses replacement datasets and models during poisoning. On the other hand, Perfect-knowledge attacks simulate a more real attack, as the adversary knows everything about the targeted system.

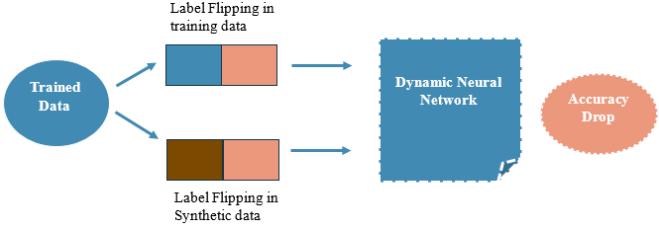

Fig. 1. Data poisoning attack.

Data poisoning is a limited knowledge attack where the adversary is unaware of the model's architecture. This attack adversely affects the system's performance. Typically, we assume that the adversary can introduce a certain percentage of fake data points into the training dataset by altering the attack classification column. The adversary can trick the system by creating data points that are difficult to detect with outlier detection techniques and data pre-filtering. In this attack, the attacker selects a percentage of instances at random from the training dataset and changes their labels to a class different from the original one [30]. This attack only demands that the adversary must be well aware of the training labels. Another type of attack is the zero-day attacks which refer to the attacks that have yet not been detected or for which the patches have not been released. Hence, they have zero days to be rectified. Before you begin to format your paper, first write and save the content as a separate text file. Complete all content and organizational editing before formatting. Please note sections A-D below for more information on proofreading, spelling and grammar.

## IV. PROPOSED IDS FRAMEWORK

This section describes our proposed IDS framework and illustrates the methodologies to generate the data at each stage to build a detection model from labeled network data. Fig. 2, defines the high-level overview of the proposed framework.

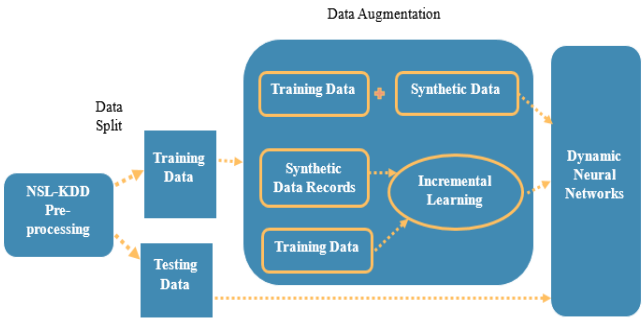

Fig. 2. Proposed IDS Framework

### A. Dataset description and pre-processing

The NSL-KDD dataset is a refined version of the KDD Cup dataset [31], widely used for benchmarking intrusion detection systems (IDS). NSL-KDD includes normal traffic and simulated attack traffic, making it highly suitable for anomaly-based intrusion detection methods. The dataset is structured into four main categories of attacks: DoS (Denial of Service) which attempts to shut down a machine or network, making it unavailable to users; Probe which includes scanning and probing to gather information about the target system; R2L (Remote to Local) in which unauthorized access from a remote machine to a local system and U2R (User to Root) which attempts to gain unauthorized root access on the target system.

The existing literature has pointed out that the extraction of relevant features can improve the model's detection accuracy, error margin, and performance. After collecting the labeled network intrusion data, the dataset is cleaned. The first and foremost step is to exclude the irrelevant columns from the dataset: the constant columns, columns whose data is independent of the result. This step must be taken based on domain knowledge. Once the data is clean, we can proceed with relevant feature extraction. Based on the data organization, we can try correlation analysis or dimension reduction techniques like PCA and TSNE on the dataset [32]. This analysis would draw the correlation between every column and thus we could exclude the highly correlated columns. We also extracted the features for four kinds of attacks separately. After cleaning, encoding, and dimension reduction of the dataset, we are left with 13 columns for each type of attack as well as for the original dataset as a whole.

### B. Dataset augmentation

Synthetic Data is artificial data generated by using various methodologies that agree with the statistical properties of the original data. The utility of synthetic data is to generate fake data, increase training dataset size, and test purposes. Various algorithms are employed to generate synthetic data, such as generating data according to a distribution, using a variational autoencoder, fitting data to distribution, and generative an adversarial network. This paper utilizes a conditional generative adversarial network such as CTGAN to generate synthetic data. This GAN readily avoids data imbalance issues as we can conditionally generate imbalanced samples from it as well. After cleaning the dataset, it is passed through the CTGAN synthesizer to generate a 1000-row dataset for the entire dataset and also for the attacks that have a lesser number of records like R2L. The data follows a similar distribution and statistical properties as the original dataset. Subsequently, for every generated instance, the probability of occurrence of the instance from each mode is computed. Finally, the calculated probability density for that instance is sampled from one mode and used to normalize the values of the instance. The original dataset was also corrupted by 20% to simulate the data poisoning effect. We estimated the data corruption using a corruption matrix that contains the number of corrupted labels for each type of attack and predicted the accuracy drop and loss using gold loss correction [19]. We have also tried to actualize the zero-day attacks because of the diversity of the testing dataset which harbors new attacks, unseen by the training dataset. The NSL-

KDD dataset is divided into training and testing in such a way that there are some new attacks in the testing dataset which are not there in the training dataset. It helped in simulating the zero-day attacks.

CTGAN architecture (as shown in Fig 3), proposes two techniques to enhance tabular data generation: mode-specific normalization and conditional training-by-sampling.

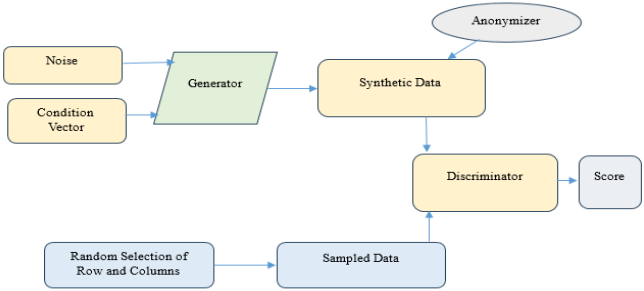

Fig. 3. CT-GAN architecture

Mode-specific normalization uses variational Gaussian mixture (VGM) modeling to determine the number of modes for each column and normalize the values accordingly [33]. Conditional training-by-sampling approach is used to condition the categorical columns. Categorical columns are strongly correlated with numerical columns hence, the data is randomly conditioned on each categorical column for each sample which thus generates the other features. This also acts as the filtering condition for sampling from the actual data distribution. The training data for the IDS is thus constructed by combining the synthetic data generated from CTGAN and preprocessed data. This synthetically generated dataset thus, overcomes the data scarcity and data imbalance problems associated with the preprocessed dataset.

*C. Incremental Learning*

Incremental learning is the machine learning approach where data is updated gradually, without the need to retrain from scratch for the new data. Most of the current applications are constrained by classical batch settings where data is provided before the training process. Hence meta-parameter optimization and model selection are based on the dataset provided and cannot be extended with data growth. Systems may employ incremental learning to restrict learning algorithms' memory and time consumption of training with new data at a manageable level because the model must make predictions following the whole data. There are two possible case scenarios for incremental learning: concept drift and true incremental learning. In the case of concept drift, old examples may be misleading, as they are outdated and different from the concept one is trying to learn. While, in true incremental learning, the model contains the same information about the target concept. There are two approaches for implementing continual learning: the instance-incremental method and the batch-incremental method. In the batch-incremental method, the incremental learning algorithm is executed after the size of new data is equal to the batch size. In contrast, in the instance-incremental method, there is no such constraint on the size of new data to trigger the incremental learning method. The instance-incremental methods perform similarly to their equivalent batch-learning implementation while using fewer resources [14].

Although IDS is utilized to detect a wide range of cyber-attacks and threats, the practice of using stale data is not fruitful. The framework model is equipped with batch-incremental learning methods to extend the knowledge of the current model with new attacks and threats. For this scenario, synthetic data is generated both for the known attacks and the hypothetical attacks. For generating the hypothetical attacks, we use conditional sampling post synthetically i.e. tampering of synthetic data previously generated to change the distribution of the data and then running the CTGAN again on it. We used batch-incremental learning methods for this purpose. The batch-learning method is used to update the existing model when the input size is equal to the batch size. This learning method also demands a need for additional memory component in the system. The extension of the memory component is acceptable since the cost of computational resources is more than the memory resources. The framework also governs attribute-based incremental tasks for training the dynamic neural network through a dynamic imputation and preprocessing module along with the retraining section for adversarial robustness. The attribute-based incremental learning task ensures that the model utilizes the additional or minimal features without reconfiguring the system.

*D. Dynamic Neural Networks*

Dynamic Neural Networks are adaptive neural networks that change their architecture based on the input sample or inference. They encompass an attention mechanism to focus on the relevant part of the information. The proposed IDS framework employs a dynamic neural network that demonstrates a functional behavior where it adjusts its layers according to the input and conditionally incorporates and dynamically imputes and preprocesses various input structures in case of missing and unstructured values. The dynamic neural network consists of a feed-forward neural network where backpropagation is managed by Pytorch's, Automatic gradient module that manages the automatic calculation of gradients. The module list of the dynamic neural network consists of the flexible hidden layers that change based on the input sample distribution and the model capacity. The maximum size of hidden layers is kept at 50, however, it is modifiable and the batch size for experimentation was kept at a threshold of 10. If it exceeds 10, then the dynamism of the architecture occurs. The addition of dynamic preprocessing and imputation modules makes it more robust towards model convergence issues and programming failures.

## V. RESULTS AND DISCUSSIONS

The platform used was Google Collaboratory and H20 Web UI. NSL KDD dataset acts as the gold standard dataset for this framework. The dataset was thus trained and tested for different case scenarios. Initially, it was tested using the entire training and testing dataset but then eventually divided into 4 subsets DOS, Probe, R2L, and U2R. The initial dataset implementation performed multi-class classification of 23 known attacks and simulated the zero-day attack and data poisoning attack for known and unknown attacks in the testing dataset while the subsets focused on binary classification between normal and other attacks. The classes mentioned above were assigned using a dictionary mapping. Furthermore, batch-level incremental learning was demonstrated for binary classification. The dataset is tested with statistically generated fake data in multiple scenarios as shown in Table I and Table II. We have chosen a comprehensive set of five classifiers each for six different case scenarios, of which dynamic neural networks clearly emerged as the lead. The dynamic neural network is indeed robust against the poisoning and zero-day attacks with an accuracy rate of 82.7 and 53.7 respectively.

We have used label flipping strategy for generating the poisoned data. Algorithmically, we changed the labels of the label column of the dataset. We have corrupted about 20% of the data in the dataset to showcase a moderate attack. The accuracy drop was slightly less than the other models on the proposed IDS framework as compared to a significant accuracy drop in other models because of the retraining module of the proposed IDS. The incremental learning time for 50 epochs with 10 iterations for each epoch is 43.63 seconds and an accuracy rate of 82.23% while retraining ensures an accuracy of 80.13% with a run rate of 63.67 seconds with the dynamic neural network. The dynamic imputation of random missing values in the new instances helps prevent runtime errors.

TABLE I. MULTICLASS CLASSIFICATION RESULTS FOR VARIOUS MODELS

| Case Scenarios | Metrics | K Nearest Neighbor (KNN) | Random Forest | Gradient Boosting Machine (GBM) | Multi-Layer Perceptron | Proposed IDS Framework |
|---|---|---|---|---|---|---|
| **Original NSL-KDD dataset** | **Accuracy** **Log-Loss** **F1-score** | 89.57 1.0056 0.841 | 98.61 0.0068 0.997 | 98.69 0.1557 0.998 | 93.21 1.9983 0.985 | 99.92 0.0054 0.996 |
| **NSL-KDD test data with zero-day attack** | **Accuracy** **Log-Loss** **F1-score** | 53.2 3.525 0.585 | 48.6 3.256 0.738 | 48.23 4.131 0.512 | 50.6 4.020 0.543 | 82.70 1.005 0.885 |
| **Poisoned NSL-KDD Training dataset** | **Accuracy** **Log-Loss** **F1-score** | 31.0 2.855 0.367 | 41.6 2.080 0.434 | 37.4 3.994 0.418 | 44.7 2.759 0.492 | 53.7 2.017 0.597 |
| **Augmented NSL-KDD dataset with CTGAN data** | **Accuracy** **Log-Loss** **F1-score** | 96.5 0.0092 0.997 | 98.2 0.0073 0.992 | 98.78 0.1532 0.996 | 95.22 1.0056 0.994 | 99.93 0.0049 0.997 |
| **NSL KDD dataset predictions with retraining for new instances of data** | **Accuracy** **Log-Loss** **F1-score** **Learning time (sec)** | NA | 83.13 0.0104 0.887 73.4 | 81.82 1.2583 0.873 75.67 | 78.67 1.967 0.853 72.39 | 80.13 1.348 0.867 63.67 |
| **NSL KDD dataset predictions with incremental learning and dynamic imputation** | **Accuracy** **Log-Loss** **F1-score** **Learning time (sec)** | Incremental Learning is not fully attainable | Incremental Learning is not fully attainable | Incremental Learning is not fully attainable | 80.66 1.064 0.859 45 | **82.23** **1.032** **0.8864** **43.63** |

TABLE II. BINARY CLASSIFICATION RESULTS FOR VARIOUS MODELS

| Attacks | Metrics | KNN | Random Forest | SVM | Ensemble | Proposed IDS Framework |
|---|---|---|---|---|---|---|
| **DOS** | Accuracy Precision Recall F1-score | 0.997 0.996 0.996 0.996 | 0.997 0.999 0.996 0.997 | 0.993 0.991 0.994 0.992 | 0.998 0.998 0.997 0.997 | 0.998 0.997 0.997 0.997 |
| PROBE | Accuracy | 0.990 | 0.996 | 0.984 | 0.992 | 0.997 |
| | Precision | 0.986 | 0.997 | 0.969 | 0.987 | 0.998 |
| | Recall | 0.985 | 0.993 | 0.983 | 0.989 | 0.996 |
| | F1-score | 0.985 | 0.994 | 0.976 | 0.987 | 0.996 |
| R2L | Accuracy | 0.967 | 0.997 | 0.967 | 0.972 | 0.997 |
| | Precision | 0.953 | 0.970 | 0.948 | 0.957 | 0.989 |
| | Recall | 0.954 | 0.874 | 0.962 | 0.962 | 0.988 |
| | F1-score | 0.953 | 0.905 | 0.955 | 0.960 | 0.978 |
| U2R | Accuracy | 0.997 | 0.980 | 0.996 | 0.997 | 0.997 |
| | Precision | 0.931 | 0.975 | 0.910 | 0.942 | 0.969 |
| | Recall | 0.850 | 0.969 | 0.829 | 0.866 | 0.970 |
| | F1-score | 0.878 | 0.972 | 0.848 | 0.896 | 0.979 |

## VI. Conclusion

The paper highlights a dynamic neural network-based IDS framework that is trained with augmented data and equipped with incremental learning methodologies. The augmented data generation mechanism using conditional generative adversarial network (CTGAN) has handled the data scarcity and data imbalance issues well along with attack simulation. This framework introduces an incremental learning algorithm and retraining module in the dynamic neural network with flexible hidden layers and adjustable batch size to extend the existing knowledge of the learning model with the latest cyber threats and cyber-attacks. Extensive experimental validations have been thus conducted on the NSL KDD dataset where it has been compared with the known classifiers. The data poisoning attack section demonstrates that a classifier trained with augmented data and preprocessed data can detect the data's poisoning with higher accuracy than other models. Thus, the proposed model significantly enhances accuracy when baseline classifiers fail to converge. This adversarial training methodology ensures that the IDS is more adaptable to dynamic network environments, identifying novel intrusion patterns more accurately. This approach ensures robust protection for the highly dynamic, ultra-reliable, and low-latency communication requirements of 5G/6G, safeguarding critical data flows and network performance. Further improvements such as the addition of flexible dropout layers with regularization could mitigate the adverse effects and may increase the accuracy.